\documentclass[journal]{IEEEtran}
\ifCLASSINFOpdf
\else
\fi
\usepackage{graphics,graphicx}
\usepackage{latexsym}
\usepackage{algorithm,algorithmic}
\usepackage{array}
\usepackage{mdwmath}
\makeatletter
\newcommand*{\rom}[1]{\expandafter\@slowromancap\romannumeral #1@}
\usepackage{mdwtab}
\usepackage{eqparbox}
\usepackage{stfloats}
\usepackage{latexsym}
\usepackage{amssymb}
\usepackage{amsmath}
\usepackage{graphicx}
\usepackage{csquotes}
\usepackage{balance}
\usepackage[justification=centering]{caption}
\DeclareGraphicsExtensions{.eps,.pdf,.gif,.png,.jpg}
\usepackage{color}
\usepackage{hyperref}
\begin{document}
%
\title{ Wireless Network Slicing: Generalized Kelly Mechanism Based Resource Allocation}
%
%
%

\author{Yan~Kyaw~Tun,~
	Nguyen~H.~Tran,~\IEEEmembership{Senior~Member,~IEEE,~}
	Duy~Trong~Ngo,~\IEEEmembership{Member,~IEEE,~}\\
	Shashi~Raj~Pandey, Zhu~ Han,~\IEEEmembership{Fellow,~IEEE,~}
	and~Choong~Seon~Hong,~\IEEEmembership{Senior~Member,~IEEE}
	
\thanks{Manuscript received October 10, 2018; revised June 16, 2019; accepted June 29, 2019; data of current version . .; date of publication ... This work was supported by the National Research Foundation of Korea (NRF) grant funded by the Korea government (MSIT) (NRF-2017R1A2A2A05000995).  *Dr. CS Hong is the corresponding author.}
\thanks{Yan Kyaw Tun, Shashi Raj Pandey, and Choong Seon Hong  are with the Department of Computer Science and Engineering, Kyung Hee University,  Yongin-si, Gyeonggi-do 17104, Rep. of Korea, e-mail:{\{ykyawtun7, shashiraj, cshong\}@khu.ac.kr}.}
\thanks{Nguyen H. Tran is with the School of Information Technologies, The University of Sydney, NSW 2006, Australia, email{\{nguyen.tran\}@sydney.edu.au}.}
\thanks{Duy Trong Ngo is with the School of Electrical Engineering and Computing, The University of Newcastle, Callaghan, NSW 2308, Australia, email{\{duy.ngo\}@newcastle.edu.au}.}
\thanks{Zhu Han is with the Electrical and Computer Engineering Department,
    	University of Houston, Houston, TX 77004, and the Department of
    	Computer Science and Engineering, Kyung Hee University, Yongin-si,
    	Gyeonggi-do 17104,  Rep. of Korea, email{\{zhan2\}@uh.edu}.}}

\maketitle

  \begin{abstract}
Wireless network slicing (i.e., network virtualization) is one of the potential technologies for addressing the issue of rapidly growing demand in mobile data services related to 5G cellular networks. It logically decouples the current cellular networks into two entities; infrastructure providers (InPs) and mobile virtual network operators (MVNOs). The resources of base stations (e.g., resource blocks, transmission power, antennas) which are owned by the InP are shared to multiple MVNOs who need resources for their mobile users. Specifically, the physical resources of an InP are abstracted into multiple isolated network slices, which are then allocated to MVNO's mobile users. In this paper, two-level allocation problem in network slicing is examined, whilst enabling efficient resource utilization, inter-slice isolation (i.e., no interference amongst slices), and intra-slice isolation (i.e., no interference between users in the same slice). A generalized Kelly mechanism (GKM) is also designed, based on which the upper level of the resource allocation issue (i.e., between the InP and MVNOs) is addressed. The benefit of using such a resource bidding and allocation framework is that the seller (InP) does not need to know the true valuation of the bidders (MVNOs). For solving the lower level of resource allocation issue (i.e., between MVNOs and their mobile users), the optimal resource allocation is derived from each MVNO to its mobile users by using KKT conditions. Then, bandwidth resources are allocated to the users of MVNOs. Finally, the results of simulation are presented to verify the theoretical analysis of our proposed two-level resource allocation problem in wireless network slicing.     
  \end{abstract}
  

\section{Introduction}
\label{sec:pro1}
\IEEEPARstart{N}{owadays}, wireless networks have faced with an explosive growth of mobile data traffic because of the dramatic increase in the use of mobile devices, and consequently, data greedy applications. To address the ever growing network traffic, in recent years, wireless network slicing has been a central topic of research. Wireless network slicing decouples mobile network operators (MNOs) in the current wireless network into two bodies: InPs and MVNOs. The physical wireless network including physical infrastructure such as base stations, cell sites, radio towers, antennas, physical resource blocks (RBs), backhaul, core network, transmission networks, transmission power, etc., are owned and operated by an InP. Physical resources from multiple InPs are leased by the MVNOs to create their own virtual networks for delivering particular services such as VoIP, live streaming, video conferencing, and video telephony, to their network users. By enabling the sharing of physical resources, wireless network slicing enables effective reduction in operational expenditures (OPEX) and capital expenditures (CAPEX) of mobile network operators (MNOs). It also enables a flexible network operation by facilitating the coexistence of multiple MVNOs on a shared infrastructure \cite{liang2015wireless}.

Though network slicing is the potential technology for future mobile networks, there remains several challenging issues to address. Among them, one important issue is how to efficiently slice and split radio resources (i.e., bandwidth or physical RBs) into multiple slices for MVNOs who must meet the dynamic demands of their mobile end users, whilst ensuring the key requirements of \emph{inter-slice} and \emph{intra-slice} isolation \cite{kokku2012nvs}, \cite{haider2009challenges}. In this regards, as defined in 5G architecture proposed by \cite{alliance20155g}, virtualization of network functions relies on network function virtualization (NFV) and software defined network (SDN) technologies. Specifically, NFV enables the abstraction of the resources and facilitates in sharing them among multiple tenants for future network services \cite{ordonez2017network}. Here, the virtualization layer, referred to as a hypervisor, enables an agile network environment which is managed by the SDN-based open standard application programming interface (API). A number of SDN controllers have been developed to enable a flexible and programmable radio access network (RAN), namely the SD-RAN platform \cite{chen2014softmobile}, \cite{foukas2016flexran}, \cite{wu2014pran}. In \cite{kokku2012nvs}, the authors designed a virtualization substrate, in particular a flow scheduler, and implemented it to meet the key requirements of efficient resource utilization, customization, and isolation in wireless resource virtualization. In \cite{foukas2016flexran}, the authors designed the controller, namely FlexRAN that uses an agent API which transparently communicates to UEs. The control protocols of such controller can make scheduling decisions such as resource block allocation. With the software enabler, the eNodeB only has to handle the data plane and the operations such as obtaining and setting the configurations, applying the scheduling decisions, maintaining the flow and so on are abstracted via the control plane with the help functions provided by the APIs. Similar to these described virtualized wireless network architecture based on SDN in \cite{kitindi2017wireless}, \cite{nfv2013etsi}, \cite{addad2018towards}, \cite{zhang2018scalable}, network slicing functionality can be considered one instance of the virtual machine (VM) in our proposed system model.

Efficient resource allocation helps to improve resource utilization, ensures the quality of services for the end users and furthermore, provides energy efficiency. The resource allocation problem in wireless network slicing is more challenging when selfish agents (i.e., MVNOs) are involved. Therefore, under such scenario where the agents act greedily, it is important to design an appropriate incentive plan in order to achieve social efficiency. In this regard, to address the challenges in the efficient resource allocation in wireless network slicing, two prominent frameworks are implemented in the wireless network slicing. In the first approach, the InP acts as a central player and can directly allocate resources to mobile users of MVNOs as per the predetermined resource requirement. In the second approach, MVNOs take part in the resource scheduling to their users instead of the InP. Firstly, the InP interacts with MVNOs and allocates resources to them. Then, the MVNOs will manage the individual resource allocation (i.e., scheduling) to their own mobile users. Therefore, with the involvement of MVNOs, the resource allocation design corresponds as a two-level problem. Most of the existing research works investigated the first resource allocation design where they ignored the role of MVNOs \cite{malanchini2014generalized}, \cite{van2014dynamic}, \cite{kamel2014lte}, \cite{lu2014elastic}. Unlike existing works that only focus on maximizing network utilization, our problem formulation considers the network economics issue in wireless network slicing. It includes monetary profit to the InP in terms of efficient resource allocation strategy for multiple associated MVNOs, and the corresponding economic interactions between MVNOs and its users. In this work under wireless network slicing, we focus on the \emph{two-level resource allocation problem} to maximize the individual and the aggregate valuation of MVNOs. Here, the most important challenge is the resource allocation among MVNOs with fairness guarantee.       
  
 Under the aforementioned challenges, we design a generalized Kelly mechanism (GKM) \cite{ma2016efficient} to address the upper-level problem and make use of the Karush-Kuhn-Tucker (KKT) conditions in addressing the lower-level of the resource allocation problem. The GKM belongs to one of the auction algorithms where each agent (i.e., MVNO) can submit an individual bid for resources to the seller (i.e, InP), while an InP receives bids from different MVNOs and then allocates resources to each bidding agent (i.e., MVNO) proportionally to their bidding values \cite{yang2013price}. The Kelly mechanism (KM) \cite{kelly1997charging} is suitable for price-taking agents, i.e., agents who have no power to influence the market price of the available resources with their bidding value. That is only possible when there are a large number of agents (i.e., MVNOs) in the resource allocation auction. However, the GKM is suitable for both price-anticipating and price-taking agents. Here, the price-anticipating agent means an agent's bidding value can influence the market price of the resources. Such price anticipating agents' bidding values may lead to loss in efficiency, and the social welfare (i.e., sum of all MVNO's valuation). At that time, GKM can reduce the loss of efficiency. Note that there are several effective auction mechanisms such as the Vickrey-Clarke-Groves (VCG) \cite{nisan2007computationally} which focuses on
 the scenario where the agents bid truthfully, i.e., every agent has to submit its true valuation as a bid. However, as valuation is the private information of agents, they will not submit it to the seller. In the GKM, even if the agents do not submit their true valuations, the seller can still induce the marginal valuations of the agents. 

\subsection{Research Contributions}  
 In order to address the challenges and issues of resource allocation in wireless network slicing as mentioned above, we propose an efficient resource allocation framework by using the GKM. Summary to our main contributions is:   
 \begin{itemize}
 \item Firstly, a two-level resource allocation problem in wireless network slicing is proposed. Then, the GKM is designed to address the upper-level of the proposed resource allocation problem. In the GKM, MVNOs will submit their individual bidding values to the InP in order to request wireless resources. The InP will further allocate its physical resources to MVNOs according to their bidding values. Then, each MVNO will use that wireless resources allocated by the InP to serve its mobile users. The most important challenges of the resource allocation in the network slicing such as isolation and fairness between MVNOs are handled by the proposed problem formulation. 
\item We next perform the theoretical analysis of GKM properties such as the existence of a unique Nash equilibrium, and the optimal resource allocation to MVNOs under the Nash equilibrium. Then, we analyze the influence power of each bidder (i.e., MVNO) in the market which is the ability of the MVNO to change the market price of the resources. To control the market influence power of the bidders (i.e., MVNOs), the seller (i.e., InP) introduces the penalty value that is attached with the cost for each bidder in GKM. We further analyze the effect of this penalty value for each MVNO under the Nash equilibrium.       
\item Finally, we use KKT conditions to address the lower-level of the proposed problem (i.e., between MVNOs and their mobile end users), and provide the closed-form solution to this problem. Moreover, we also consider an incomplete information scenario in which each MVNO does not know the channel condition of its mobile users due to estimation error, or wireless channel delay. We further extend our work into multiple resources scenario where each MVNO requests multiple resources (e.g., bandwidth, power) from an InP. 
\item In simulation section, we first present the resource allocated to MVNOs under the GKM. Then, we compare the achieved valuation of each MVNOs under proposed GKM with others: Equal Sharing, traditional Kelly mechanism, and Optimal solution. The proposed scheme achieves a significant performance gain: up to $13\%$, and $9\%$ in comparison to Equal Sharing, and traditional Kelly mechanism with our proposed algorithm, respectively. We also observe that our proposed solution framework achieves near Optimal solution. Further, we also demonstrate the allocated bandwidth to each user of MVNO under KKT conditions.              
 \end{itemize} 
 
The rest of this paper is organized as follows: Section \ref{Related} summarizes related works. The system model and wireless network slicing framework are introduced in Section \ref{sec:pro3}. Section \ref{sec:pro4} presents the two-level resource allocation problem in wireless network slicing and proposes the solution mechanism. The extension of our proposed multiple resource allocation problem in wireless network slicing is presented in Section \ref{sec:pro5}. Section \ref{sec:pro6} discusses about the simulation results. Finally, Section \ref{sec:pro7} concludes the paper.
 \begin{figure}[t!]
	\centering
	\captionsetup{justification=centering}
	\includegraphics[width=3.7in]{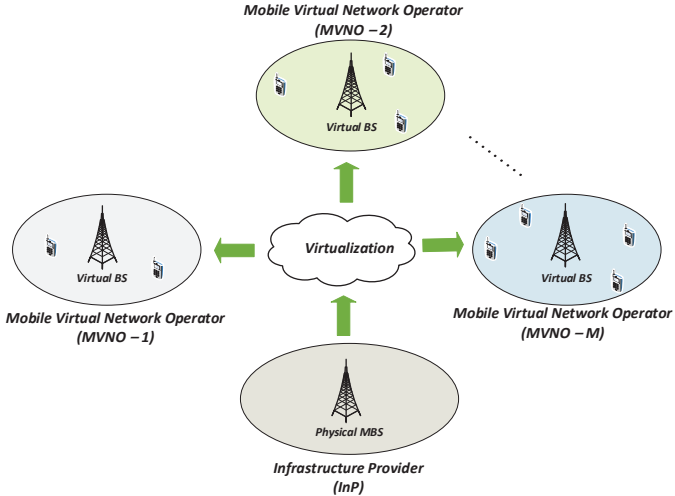}
	\caption{A model of wireless network slicing.}
	\label{fig:systemmodel}
\end{figure}
\section{Related Works}
\label{Related} 
	Resource layer, along with network slice instance layer, and service instance layer is one of the integral parts of network slicing in 5G architecture proposed by \cite{alliance20155g}. A network slice supports at least one type of service, and should be mutually isolated, manageable and support multi-tenants, multi-services \cite{galis2018network}, \cite{nakao2017end}. In this regard, the recent 3GPP R15 \cite{3gppr15} specifications and standards define tailored services such as  massive machine type communication (mMTC), ultra reliable and low latency communications (URLLC), and enhanced mobile broadband (eMBB). Therefore, a proper design of the resource allocation solution that is flexible, scalable and demand-oriented in wireless virtualization has to be done, which is the scope of this paper.
 
Proportional allocation (i.e., Kelly mechanism) in one-sided resource allocation auction was investigated in \cite{johari2004efficiency}, \cite{correa2013price}. They showed that under the assumption of price-taking agents, Kelly mechanism achieves maximum value of social welfare. In order to reduce the loss efficiency gap and the market influence power of the price-anticipating agents, \cite{ma2010resource} studied a GKM by setting a penalty value for each price-participating agent according to their bid. Moreover, the theoretical limitations of both GKM and Kelly mechanism were presented in \cite{syrgkanis2013composable}.

In \cite{fu2013stochastic}, the authors have proposed a stochastic game-based spectrum allocation in virtualized wireless networks. Although the proposed resource allocation scheme achieved higher resource utilization, MVNOs are not considered in resource allocation design. Moreover, as InP manages resources and allocates it directly to the mobile users of MVNOs in a centralized manner, the computation complexity is high.  

The work of \cite{tun2017downlink}, \cite{parsaeefard2015joint} introduced a joint resource allocation and admission control strategy for an orthogonal frequency division multiple access (OFDMA) based virtualized wireless network. Here, both resource-based and rate-based MVNOs were considered, and a joint optimization problem for power and resource allocation was formulated for maximizing the overall sum rate of the corresponding MVNOs. But the significance of MVNOs was ignored in \cite{tun2017downlink}, \cite{parsaeefard2015joint} and the user scheduling was instead performed by the base stations of the InP. The dynamic resource management in wireless virtualized networks was proposed in \cite{zaki2010lte}. The developed dynamic resource sharing approach can result in higher resource utilization and better system efficiency.

 In \cite{bashar2009admission}, the authors introduced an auction game model for the users to bid for radio resources. Moreover, auction mechanism based power allocation in the LTE air interface virtualization was proposed in \cite{fan2015game}. The authors of \cite{zaki2010novel} proposed an LTE framework with an added entity called \enquote{hypervisor} at a base station. The hypervisor enables sharing of RBs among the MVNOs without interfering with each other. In \cite{zhu2016virtualization}, a combination of wireless network virtualization and massive MIMO was considered. Then, the authors formulated a resource (i.e., bandwidth, power, antennas) allocation problem as a hierarchical structure and implemented a combinatorial VCG auction mechanism for solution. In most of the existing works, the responsibility of MVNOs was missing and they did not consider economic models of wireless network slicing. In our work, we consider both two-level resource allocation problem and economic model of wireless network slicing. Moreover, we highlight the responsibility of MVNOs in the resource allocation in wireless network slicing.
\begin{table}[t]
    \centering
	\caption{Summary of Notations.}
	 \label{tab:table1}
	\begin{tabular}{ll}
		\hline
		Notation & Definition\\
		\hline
		$\mathcal{M}$ & Set of MVNOs,  $|\mathcal{M}|= M$\\
	
		$\mathcal{S}_m$ & Set of mobile users of MVNO $m \in \mathcal{M}$, $|\mathcal{S}_m|= S_m$\\
		
		$R$  & Total bandwidth capacity of an InP \\ 
		
		$r_m$ & Bandwidth allocated to MVNO $m \in \mathcal{M}$ \\
		
		$r_m(\mathbf{b})$ & Bandwidth allocated to MVNO $m \in \mathcal{M}$ depends on \\
		& the bidding vector $\mathbf{b}$ of MVNOs\\ 
		$\mathbf{b}$ & The vector of bidding values of MVNOs \\
		$b_m$ & Bidding value of MVNO $m \in \mathcal{M}$\\
		$B$ & Sum of bidding values of all MVNOs \\
		$v_m(r_m(\mathbf{b}))$ & Valuation of MVNO $m$ depends on the \\ & allocated resource $r_m(\mathbf{b})$   \\
		$c_m(\mathbf{b})$ & Cost function of MVNO $m \in \mathcal{M}$  \\
		$q_m$ & Penalty value of MVNO $m \in \mathcal{M}$ \\
		$\mathbf{q}$ & The vector of penalty values of MVNOs \\
		$v'(r_m(\mathbf{b}))$  & Marginal valuation of MVNO $m \in \mathcal{M}$\\
		$\mu_m$ & Market influence power of the MVNO $m \in \mathcal{M}$ \\
		$\beta$ & The virtual price of bandwidth (per Hertz)\\
 		$x_s^m$ & The allocated resource to the user $s \in \mathcal{S}_m$ of MVNO \\
 		& $m \in \mathcal{M}$  \\
 		$\mathbf{E}$ & The resource competition matrix \\
 		$\mathbf{e}_\mathbf{m}$ & The vector of the resource allocated to MVNO $m \in \mathcal{M}$ \\
 		$\mathbf{Q}$ & The penalty matrix\\
 		$\mathbf{B}$ & The bidding matrix \\
		\hline
	\end{tabular}
\end{table}

\section{System Model}
\label{sec:pro3}

A wireless network slicing where a single InP having a physical macro base station (MBS), and a set of mobile virtual network operators, $ \mathcal{M} = \{1,2,\ldots,M\}$ that provides particular mobile services to their users is shown in Fig.~\ref{fig:systemmodel}. Specifically, in this work, we consider 4G architecture for general virtualization, similar with the works in \cite{kamel2014lte}, \cite{kokku2012nvs}. The MBS is operating on the total bandwidth of $ R $ and each MVNO $m \in \mathcal{M}$ provides services to the users $ \mathcal{S}_m =\{1,2,\ldots,S_m \}$. A fraction of the total bandwidth $R$ allocated to each MVNO $m \in \mathcal{M}$ is defined as $r_m$\footnote{ Considering the 3GPP specification and standards for 4G architecture \cite{3gpp}, the resource allocation problem considers resource block (RB) as the minimum allocation unit. However, to make the problem tractable, we use the continuous form of resource as `bandwidth', similar to the works in  \cite{uzawa2019first}, \cite{chen2018efficient}, \cite{wu2018common}, to solve the problem.}. Here, a hypervisor is deployed by the InP at the MBS to slice its physical resources for leasing among multiple MVNOs. A central question is how the InP will schedule its wireless bandwidth amongst multiple MVNOs to give services to their mobile users. Because the InP cannot access the information of users such as QoS requirements and channel conditions. Therefore, a possible solution is to allocate bandwidth to MVNOs first, and afterwards each MVNO allocates the wireless bandwidth to its users. This approach is regarded as a two-level solution approach.   

\begin{figure}[t]
\centering
\includegraphics[width=3.5in]{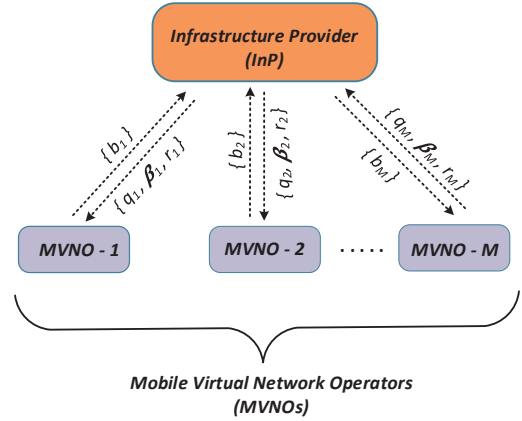}
\caption{Generalized Kelly Mechanism.}
\label{fig:GKM}
\end{figure}

In this work, the resource (i.e., bandwidth) allocation problem in wireless network slicing is decomposed into two levels. In the upper level, as shown in Fig.~\ref{fig:GKM}, the InP decides how to efficiently allocate bandwidth to multiple MVNOs and which aims to maximize the social welfare (i.e., aggregate valuation of MVNOs). In the lower level, each MVNO manages resource scheduling to its mobile users by considering its own utility. We formulate the upper-level problem as an auction-based resource allocation problem for which the GKM is proposed for solution. Each MVNO $m \in \mathcal{M}$ will report
 its own bidding value $b_m$ $(0\leq b_m < \infty )$ to the InP in each resource allocation round. Depending on the bidding values of all MVNOs, they will receive a proper allocation of bandwidth form the InP. The bandwidth allocation among MVNOs will be straightforward when the InP knows the characteristics (i.e., valuation) of the MVNOs. However, the valuation function is the private information of each MVNO and it is related with the dynamic channel conditions of its users. After that, each MVNO will assign the bandwidth to its users as per their QoS requirements.

\section{Two-Level Resource Allocation in Wireless Network Slicing}
\label{sec:pro4}
In this work, the resource allocation problem in the wireless network slicing can be decoupled into two levels: 1) resource allocation between InP and MVNOs, and 2) resource allocation from MVNO to its mobile users. 

\subsection{Upper-level Problem} 

Depending on the number of users and their QoS requirement, each MVNO decides the required wireless bandwidth. Let us define the valuation function $v_m(r_m(\textbf{b})), \mathbf{b} = \{b_1, b_2, \ldots, b_M\}$ is the vector of the bidding value, as the satisfaction of the MVNO $ m \in \mathcal{M}$.     \\

       \textbf{Assumption 1} : The valuation function $v_m(r_m(\textbf{b}))$ is strictly increasing, concave and continuous over the domain $r_m>0$ .
           
This assumption is widely used for utility or valuation functions in communication networks \cite{ma2016efficient}, \cite{luong2018applications}, \cite{jin2018auction}. Here, the InP will allocate its fraction of resource (bandwidth) to each MVNO according to the reported bidding value of MVNOs. It means that InP will allocate the largest ratio of bandwidth to the MVNO with the highest bidding value. Thus, the GKM framework \cite{ma2016efficient} can be used to express the interaction of InP and MVNOs, where the objective is to maximize the aggregate valuation of MVNOs. Therefore, the wireless bandwidth allocation to the MVNOs from the InP in a virtualized network is formulated as follows: 
            \begin{align}
            {\max}\qquad
             & \sum_{ m\in \mathcal{M} } v_m(r_m(\textbf{b})) \\
            \text{s.t.} \qquad
             & r_m(\textbf{b}) \cap r_n(\textbf{b}) = \emptyset,\text{for $m\neq n$}, \text{and}\ m,n \in \mathcal{M},  \\
             & \sum_{m=1}^{M} r_m(\textbf{b}) \leq R, \\
             \text{var.} \qquad 
             & r_m(\textbf{b})\geq 0, \ \ \forall m\in \mathcal{M},
            \end{align}  	 
which considers the optimal division of the total bandwidth $ R $ of the InP under the GKM. Constraint (2) ensures the isolation between different MVNOs. As the MBS has limited amount of bandwidth, constraint (3) guarantees the allocated bandwidth of all MVNOs not exceed the total bandwidth of the MBS and (4) ensures that the resource allocated to each MVNO must be positive value. 	

Solving problem (1)-(4) is possible once the valuation of MVNOs is known at the InP. However, the valuation is the private value of each MVNO. Therefore, the MVNOs will not share these information to the InP so as to maximize their own utilities with allocation of bandwidth. Upon submission of the bidding value $b_m$, each MVNO will receive the fraction of the total bandwidth of MBS $r_m(\textbf{b})$ accordingly. Let $ \textbf{r} =\{r_1,r_2,\ldots,r_M\}$ be the resource allocation vector which is determined by the proportional allocation \cite{ma2016efficient} as follows:  
	\begin{equation}
	r_m(\textbf{b}) = \frac{b_m}{\sum_{m=1}^{M} b_m} R ,\ \ \ \ \ \    \forall m \in \mathcal{M}, 
	\end{equation}
where $\sum_{m=1}^{M} b_m = B$ is the total bidding value at the InP. Here, the proposed resource allocation scheme guarantees the fairness among MVNOs [see Remark 1]. Each MVNO $m \in \mathcal{M}$ has the cost function $c_m(\textbf{b}) = q_m b_m$ which depends on the bidding value $b_m$ and $q_m$, the penalty parameter which is varying according to the bidding value. Then, the payoff function of the MVNO $m$ is defined as   
      \begin{equation} 
      \begin{split}
      u_m(r_m(\textbf{b})) &= v_m(r_m(\textbf{b})) - c_m(\textbf{b})  \\
       & = v_m(r_m(\textbf{b})) - q_m b_m,   \ \ \ \ \ \ \forall m\in \mathcal{M},
      \end{split}
      \end{equation} 
	 where $v_m(r_m(\textbf{b}))$ is the valuation of MVNO $m$ with allocated resource $r_m$ based on bidding value $b_m$ and the penalty vector of all MVNOs with the bidding value is $ \textbf{q} = \{q_1,q_2,\ldots,q_M\}$.\\ 
	 
	 \textbf{Remark 1.} \textit{The proportional allocation scheme can maintain fairness amongst competing MVNOs. This is because the allocation of resources is based upon the proportion of bidding values of each MVNO, i.e., bidding higher means getting more resource.}\\
	 
	 \textbf{Proposition 1}: The optimal bidding of each MVNO $m \in \mathcal{M}$ is 
	 \begin{equation}
	 b_m = \frac{1}{q_m}r_m(\textbf{b}) v'(r_m(\textbf{b}))(1- \mu_m), \ \ \  \forall m \in \mathcal{M},
	 \end{equation} 
	 where $\mu_m$ is the market influence power of the MVNO $m \in \mathcal{M}$ to be explained later in this section.
	 
	 \begin{IEEEproof}
	 See Appendix A. 
	 \end{IEEEproof}
	 
	 \textbf{Proposition 2}: The unit penalty parameter for each MVNO is
	 \begin{equation}
	 q_m = \frac{1}{\beta}v'_m(r_m(\textbf{b}))\left(1-\frac{r_m(\textbf{b})}{R}\right), \ \ \ \ \ \ \ \ \ \ \forall m\in \mathcal{M}.
	 \end{equation}
	 \begin{IEEEproof}
	 See Appendix B.
	 \end{IEEEproof}

However, the penalty parameter for each MVNO $m \in \mathcal{M}$ depends on its valuation. As the valuation of each MVNO is its private information, it will not reveal it to the InP. In such a case, one can employ an iterative algorithm that allows the InP to approximate the penalty for each MVNO from the information of the previous iteration. Therefore, the penalty for MVNO $m \in \mathcal{M}$ at the $k^{th}$ iteration is as follows \cite{yang2013price}: 
	 \begin{equation}
	 q_m^k = q_m^{k-1} + \left(\frac{R - (r_m(\textbf{b}))^{k-1}}{M-1} - \frac{Rq_m^{k-1}}{\sum_{m=1}^{M}q_m^{k-1}}\right), 
	 \forall m \in \mathcal{M}.       
	 \end{equation}
	 \begin{IEEEproof}
	 	 See Appendix C.
	 \end{IEEEproof}
	   
	  Without the penalty parameter $q_m , \forall m \in \mathcal{M}$, each MVNO will try to obtain a large proportion of wireless bandwidth resource from the InP by bidding higher, in fact, as much as possible. For this reason, the InP imposes a control parameter, defined as the penalty vector $\mathbf{q}$ to obtain true valuations of the MVNOs and balance the bandwidth allocation between the competing MVNOs. In this regards, the InP interacts with the MVNOs as follows: 1) InP informs the penalty parameter $q_m$ to each MVNO $m$, 2) considering the penalty parameter, each MVNO submits its bidding value $b_m$ to get resources from the InP, and 3) the InP broadcasts the virtual price of the bandwidth (per Hertz). Therefore, the virtual price of bandwidth (per Hertz) is  
	  \begin{equation}
	  \beta = \frac{\sum_{m=1}^{M} b_m}{R}.
	  \end{equation}
 After knowing the virtual price for the bandwidth, each MVNO can derive the fraction of the bandwidth it receives as $r_m = \frac{b_m}{\beta}, \forall m \in \mathcal{M}$. To this end, when there are few MVNOs in the resource allocation, the bidding value of each MVNO will largely influence the virtual price. Therefore, we can observe that each MVNO is capable enough to manipulate the outcome of the resource allocation game. However, an increase in the number of MVNO will eventually eliminate such influences, i.e., the market influence power of each MVNO is low. In this regards, with an infinite number of MVNO in the resource allocation, the individual market influence ability of the MVNOs approaches to zero. Note that, in the real world, an infinite number of MVNOs is not possible. Therefore, in our formulation, we have considered the market influence power of each MVNO $m \in \mathcal{M}$ as  

\begin{equation}
\mu_m = \frac{b_m}{\sum_{m=1}^{M}b_m},  \ \forall m \in \mathcal{M}. 
\end{equation} 
From (11), we observe that the market influence power of MVNO $m \in \mathcal{M}$ is coupled with the bids of other competing MVNOs, however, these bids are their private information. In such a case, one can employ an iterative algorithm that allows the MVNO to know the approximate market influence power from the information related with the previous iterations. Thus, for the MVNO $m \in \mathcal{M}$, its market influence power at the $k^{th}$ iteration can be defined as  
\begin{equation}
\mu_m^k = 1 - \frac{b_m^{(k-1)}q_m^{(k-1)}}{(r_m(\textbf{b}))^{(k-1)}v'_m\left((r_m(\mathbf{b}))^{(k-1)}\right)}.                  
\end{equation} 
 
 In our bandwidth allocation (i.e., bandwidth competition among MVNOs) game, each MVNO will adopt a strategy $b_m$ to maximize its utility $u_m(\textbf{b})$ as 
    \begin{equation}
    u_m(b_m;\textbf{b}_{-m}, \textbf{q}) = v_m(r_m(\textbf{b}))- q_m b_m ,      \ \ \ \ \ m\in \mathcal{M}, 
    \end{equation} 
where $\textbf{b}_{-m}=[b_1, \dots, b_{m-1}, b_{m+1}, \dots, b_M]$ denotes the strategy profiles of all the other MVNOs except $m$. Then, for each MVNO, with the strategy profile $b_m^*, \forall m \in \mathcal{M}$, there exists a unique Nash equilibrium in the formulated resource competition game if the following relation is satisfied:  
 \begin{equation}
 u_m(b_m^* ; \textbf{b}_{-m}^*, \textbf{q}) \geq u_m(b_m ; \textbf{b}_{-m}^*, \textbf{q}),  \ \ \ \ \ \ \   \forall b_m\geq 0.
 \end{equation} 
 
\textbf{ Theorem 1} (Uniqueness of Nash equilibrium): When $M>1$, at least two components of $b_m>0$ and Assumption 1 holds. For any $q_m\in\textbf{q}$, there is a unique Nash equilibrium for the resource competition game with the strategy profile $b_m > 0, \forall m\in \mathcal{M}$. 

\begin{IEEEproof}
See Appendix D.
\end{IEEEproof}

In order to distinguish the equilibrium conditions of this bandwidth allocation (i.e., resource competition among MVNOs), the function $\hat{v}(r_m)$ is introduced as
\begin{align*}
\hat{v}_m(r_m(\textbf{b})) &= \frac{1}{q_m} \left(1- \frac{r_m(\textbf{b})}{R}\right)v_m(r_m(\textbf{b})) \\
& \qquad \qquad \qquad + \frac{1}{q_m R}  \int_{0}^{r_m} v_m(z) dz .     \tag{15}
\end{align*}  
The efficient bandwidth allocation to MVNOs in this resource competition among MVNOs can be explored according to the following optimization problem
  \begin{align*}
             {\max}\qquad
             & \sum_{ m\in \mathcal{M} } \hat{v}_m(r_m(\textbf{b}))  \tag{16} \\
            \text{s.t.} \qquad
             & \sum_{m=1}^{M} r_m(\textbf{b}) \leq R,    \tag{17} \\
             \text{var.} \qquad
             	& r_m(\textbf{b})\geq 0, \ \ \forall m\in \mathcal{M}.   \tag{18} 
            \end{align*}  
Let $r_m^*$ be the solution to the above optimization problem.

 \textbf{Proposition 3}: There exists a unique Nash equilibrium of the resource allocation (i.e., resource competition among MVNOs) shown in Theorem 1. Under that unique Nash equilibrium, the allocated bandwidth $r_m$ to each MVNO $m \in \mathcal{M}$ is the solution to the above optimization problem shown in (16) with constraints (17) and (18).  
	 \begin{IEEEproof}
	 See Appendix E.
	 \end{IEEEproof}
\vspace{-0.3in} 
\subsection{Lower-level Problem}

In the lower-level, each MVNO aims at maximizing its valuation by allocating the obtained bandwidth from the InP. The valuation of each MVNO $ m \in \mathcal{M}$ is the sum of the logarithmic function of data rate of its users. Therefore, the valuation of an MVNO $m\in\mathcal{M}$ can be defined as: 
            \begin{align*}
	         v_m(r_m)= \qquad
             &{\max}  \sum_{s=1}^{S_m} \log\left(x_s^m r_m \log_2\left(1+ \frac{p_sh_s}{N_0}\right) + 1 \right)    \tag{19} \\ 
            \text{s.t.} \qquad
             & \sum_{s=1}^{S_m} x_s^m  \leq 1, \forall m \in \mathcal{M},   \tag{20}  \label{tag:20} \\
             \text{var.} \qquad
             &x_s^m \in [0,1],\forall s \in \mathcal{S}_m, \forall m \in \mathcal{M},   \tag{21}   \label{tag:21}    
            \end{align*}
where $p_s$ is the downlink transmitted power of the BS to a mobile user $s$. Note that we assume fixed power allocation per bandwidth (Hertz) in this paper. Moreover, $h_s$ is the channel gain of user $s$, $N_0$ is the noise power, and $x_s^m$ represents a fraction of bandwidth of MVNO $m$ assigned to $s$ where $x_s^m \in r_m$. (20) and (21) are the constraints for the fraction of wireless bandwidth allocated to each subscriber of the MVNO $m \in \mathcal{M}$. As constraints (20) and (21) are linear, the constraint set is affine and objective function (19) is concave. Therefore, the valuation function $v_m(r_m(\textbf{b}))$ of each MVNO $m \in \mathcal{M}$ satisfies Assumption 1. 
\vspace{-0.3in}                        			
\subsection{Optimal Bandwidth Allocation}
The bandwidth allocation problem in (19) is a convex problem. Thus, the optimal solutions for (19) can be obtained via Lagrangian duality \cite{boyd2004convex}. Here, the Lagrangian of (19) is  
\begin{equation*}
 				\begin{aligned} 
 				L(x_s^m,\lambda,\nu) ={} &\sum_{s=1}^{S_m} \log\left(x_s^m r_m\log_2\left(1+ \frac{p_sh_s}{N_0}\right) + 1 \right) \\
 				& + \lambda \bigg(1-\sum_{s=1}^{S_m}x_s^m\bigg), 	
 				\end{aligned}  \tag{22} 
\end{equation*}
where $\lambda \geq 0$ is the Lagrangian multiplier defined for constraint (21). By using the KKT conditions, we get the optimal bandwidth allocated to each user $s \in \mathcal{S}_m$ as
\begin{equation*}
x_s^{m*} = \frac{1}{r_m} \left( \frac{1}{|S_m|} \left[r_m + \sum_{s=1}^{S_m} \frac{1}{\alpha^{*}} \right] - \frac{1}{\alpha^{*}} \right) , \forall s \in \mathcal{S}_m.     \tag{23} 
\label{eq:resource_alloc}
\end{equation*} 
\begin{IEEEproof}
	 See Appendix F.
\end{IEEEproof}

\subsection{Lower-level Problem with Incomplete Information} 
In a practical scenario, it is hard for MVNOs to get precise information of the channels (bandwidth) because of the estimation errors, and the wireless channel delay. To address the uncertainty of the wireless channel, in this work we consider that the wireless bandwidth follows Rayleigh fading \cite{tang2017combating}. Since there is no complete information at the MVNO, we need to introduce an outage probability constraint in (19) as 
            \begin{align*}
	         v_m(r_m)= \
             &{\max}  \sum_{s=1}^{S_m} \log\left(x_s^m r_m \log_2\left(1+ \frac{p_s^m H_s^{m}}{N_0}\right) + 1 \right)    \tag{24} \\ 
            \text{s.t.} \qquad  
             &  \eqref{tag:20}-\eqref{tag:21},    \\
             & \mathsf{Prob}\left(\rho_s^{\mathsf{min}} > \log_2(1+ \frac{p_s^m h_s^m}{N_0})\right) \leq \epsilon, \ \forall s \in \mathcal{S}_m,  \tag{25}
            \end{align*}
where $\rho_s^{\mathsf{min}} = \log_2(1+\frac{p_s^m H_s^m}{N_0})$, and $\epsilon$ is a predetermined threshold on outage probability. In this work, without loss of generality, we consider that the threshold value is the same for all mobile users. Then, we can rewrite the outage probability for the QoS constraint (25) as
\begin{align*}
 &\mathsf{Prob}\left\{\rho_s^{\mathsf{min}} >  \log_2(1+ \frac{p_s^m h_s^m}{N_0}) \right\} \leq \epsilon,\\
 &\Leftrightarrow \mathsf{Prob} \left\{ \gamma_s^m \leq \frac{2^{\rho_s^{\mathsf{min}}} -1 }{p_s^m}   \right\}  \leq \epsilon, \\
 &\Leftrightarrow \rho_s^{\mathsf{min}} \leq \log_2( 1 + p_s^m F_{\gamma_s^m}^{-1}(\epsilon) ),    \tag{26}
\end{align*}
where $ F_{\gamma_s^m}(.)$ is the cumulative distribution function (CDF) of $\gamma_s^m$ for the user $s$ of the MVNO $m$ and $F^{-1}_{\gamma_s^m}(.)$ is its inverse. With Rayleigh fading, we get $ F_{\gamma_s^m}$ as: 
\begin{align*}
F_{\gamma_s^m}(a) &= \int_{0}^{a}\frac{a}{\sigma^2}e^{\frac{- (a)^2}{2 \sigma^2}} d(a) \\
&=  1 - e^{\frac{-(a)^2}{2 \sigma^2}},    \tag{27} 
\end{align*}
where $\sigma$ is the scale parameter, and $a = \frac{2^{\rho_s^{\mathsf{min}}} -1 }{p_s^m}$.

From (24), we can remove the outage probability of the QoS constraint in (26) and rewrite (19) as 
\begin{align*}
	         v_m(r_m)= \  
             &{\max}  \sum_{s=1}^{S_m} \log\left(x_s^m r_m  \log_2\left( 1 + p_s^m F_{\gamma_s^m}^{-1}(\epsilon) \right)  + 1 \right)    \tag{28} \\
             \text{s.t.} \qquad 
             &  \eqref{tag:20} \ \text{and} \  \eqref{tag:21}.                        
\end{align*}

From the KKT conditions, we get the optimal bandwidth allocation to each user $ s \in \mathcal {S}_m$ as follows
\begin{equation}
x_s^{m*}  = \frac{1}{r_m} \left( \frac{1}{|S_m|} \left[r_m + \sum_{s=1}^{S_m} \frac{1}{\nu^{*}} \right] - \frac{1}{\nu^{*}} \right) , \forall s \in \mathcal{S}_m,     \tag{29} 
\end{equation}
where $\nu^* = \log_2( 1 + p_s^m F_{\gamma_s^m}^{-1}(\epsilon))$.

			\begin{algorithm}[t!]
				\caption{\strut GKM Algorithm for Optimal Bandwidth Allocation in a Wireless Virtualized Network}
				\label{alg:profit}
				\begin{algorithmic}[1]
				    \STATE{\textbf{Initialization:} Initialize $q_m^{(0)}$, $b_m^{(0)}$, $\mu_m \leftarrow 0$, $\forall m \in \mathcal{M}$; }  
				    \STATE{\textbf{Set} $k \leftarrow$ 1} 
				    \STATE{Each MVNO $m \in \mathcal{M}$ estimates the market influence power $\mu_m^{(k)}$  according to (12);}
				    \STATE{The InP calculates the penalty for each MVNO $q_m^{(k)}$ by (9) and then informs each MVNO $ m \in \mathcal M $}
					\STATE{ Each MVNO $m \in \mathcal{M}$ updates the bidding value $b_m^{(k)}$ by (7) and then submits to the InP who then sends the virtual price $\beta^{(k)}$ in (10) to all MVNOs;}
				    
				    \FOR{$m=1$ to $M$}
				        \STATE{Based on (5), the InP calculates the amount of bandwidth $r_m^{(k)}(\textbf{b})$ to be allocated to MVNO $m \in \mathcal{M}$;}
				    \ENDFOR
				\STATE{The InP distributes $r_m^{(k)}(\textbf{b})$ to each MVNO $m \in \mathcal{M}$;}	
				\FOR{$m=1$ to $M$}
				\FOR{$s=1$ to $S_m$}
				   \STATE{Each MVNO $m$ allocates optimal bandwidth $x_s^m$ to its mobile users based on (27);}
				\ENDFOR
				\ENDFOR 
				\STATE{Each MVNO $m \in \mathcal{M}$ calculates $v_m^{(k)}(r_m)$;}
				\STATE{\textbf{Increment:} $k \leftarrow k+1;$}
				\STATE{\textbf{Repeat} lines 3 to 15 until convergence.}
				\end{algorithmic}
				\label{Algorithm}
			\end{algorithm}


\section{Wireless Network Virtualization with Multiple Resources}
\label{sec:pro5}

In this section, we consider that each MVNO $m \in \mathcal{M}$ needs $C$  divisible (e.g., power, wireless bandwidth, antennas, computation capacity, storage capacity, etc.,) at the same time to service mobile users. We can model $ C \times M $ resource competition matrix $\textbf{E}$ as 
\begin{equation*} \label{multi_1}
\textbf{E} = (\textbf{e}_\textbf{1}, \textbf{e}_\textbf{2}, \dots, \textbf{e}_\textbf{M}) = \begin{pmatrix}
e_{11} & \dots & e_{1M} \\ 
\vdots & \ddots & \vdots \\

e_{C1} & \dots & e_{CM}         

\end{pmatrix}     \tag{30} 
\end{equation*}     
where $ ( e_{c1} \ \  \dots \ \  e_{cM})$ is the row vector that indicates the allocation of the resource $c \in \mathcal{C}$ of the InP among $M$ MVNOs, and $e_m$ shows the  allocation of $C$ resources to an MVNO $ m\in \mathcal{M}$.

  \textbf{Assumption 2}: The valuation function $v_m(\textbf{e}_\textbf{m})$ is concave, strictly increasing, and continuous over the domain $\textbf{e}_\textbf{m}>0$ \cite{ma2016efficient}.\\
Here, we define the social welfare maximization problem in multiple divisible resources as 
      \begin{align} \label{multi_2} 
             {\max}\qquad
             & \sum_{ m\in \mathcal{M} } v_m(\textbf{e}_\textbf{m})   \tag{31} \\
            \text{s.t.} \qquad
             & e_{cm}  \cap e_{cn} = \emptyset,\text{for $m\neq n$}, \ m,n \in \mathcal{M}, \forall c \in \mathcal{C},   \tag{32}   \\
             & \sum_{m=1}^{M} e_{cm} \leq R_c, \ \ \   \forall c \in \mathcal{C},  \tag{33}  \\
             \text{var.} \qquad
             	& e_{cm}\geq 0, \ \ \forall m\in \mathcal{M}, \forall c \in \mathcal{C},   \tag{34} 
            \end{align} 
where $ R_c$ is the maximum capacity of resource $c \in \mathcal{C}$. (32) ensures the intra-isolation among different MVNOs for all resources $c \in \mathcal{C}$ of the InP. As the resources provided by the InP are limited, (33) guarantees that the allocated resource $e_{cm}$ to all MVNOs do not exceed the total resource capacity $R_c$.

Similar to the resource competition matrix \textbf{E}, we define the penalty  matrix $\mathbf{Q}$ and the bidding values matrix $\mathbf{B}$ of MVNOs as
\begin{equation*} \label{multi_1}
\textbf{Q} =  \begin{pmatrix}
q_{11} & \dots & q_{1M} \\ 
\vdots & \ddots & \vdots \\
q_{C1} & \dots & q_{CM} 

\end{pmatrix}   
,\ 
\textbf{B} =  \begin{pmatrix}
b_{11} & \dots & b_{1M} \\ 
\vdots & \ddots & \vdots \\
b_{C1} & \dots & b_{CM} 
\end{pmatrix},      \tag{35} 
\end{equation*}     
where $q_{cm}$ represents the penalty for MVNO $m$ to bid for resource $e_{cm}$ at the InP, and $b_{cm}$ denotes the bidding value of MVNO $m$ for the divisible resource $c$ at the InP. Let us, respectively, denote by $\textbf{q}_\textbf{m}$ and $\textbf{b}_\textbf{m}$ the penalty and bidding value with regard to MVNO $m \in \mathcal{M}$. Also denote by $Q_c$ and $B_c$ the penalty and bid with regard to the resource $c \in \mathcal{C}$, respectively. The resource $c \in \mathcal{C}$ is allocated to MVNOs according to 
\begin{equation*}
 e_{cm}(B_c) = \frac{b_{cm}}{\sum_{m=1}^{M}b_{cm}}, \ \ \ \forall c \in \mathcal{C}.     \tag{36}  
\end{equation*}       

The utility function of MVNO $m \in \mathcal{M}$ is defined as 
\begin{equation*}
u_m(\textbf{B}, \textbf{Q}) = v_m(\textbf{e}_\textbf{m}(\textbf{B}))- \textbf{q}_\textbf{m}^T\textbf{b}_\textbf{m}, \  \forall m \in \mathcal{M},   \tag{37} 
\end{equation*}   
where each MVNO chooses its bidding strategy to maximize its utility defined as 
\begin{equation*}
u_m(\textbf{b}_\textbf{m}; \textbf{b}_\textbf{-m}, \textbf{Q}) = v_m(\textbf{e}_\textbf{m}(\textbf{B}))- \textbf{q}_\textbf{m}^T\textbf{b}_\textbf{m}, \  \forall m \in \mathcal{M}.   \tag{38} 
\end{equation*}
The bidding profile matrix $\textbf{B}^*$ is a Nash equilibrium for any MVNO $m \in \mathcal{M}$ if the following equation is satisfied:

\begin{equation*}
u_m(\textbf{b}_\textbf{m}^*; \textbf{b}_\textbf{-m}^*, \textbf{Q}) \geq u_m(\textbf{b}_\textbf{m}; \textbf{b}_\textbf{-m}^*, \textbf{Q}),  \ \forall \textbf{b}_\textbf{m}\geq 0.  \tag{39} 
\end{equation*} 
The proof of the existence of a unique Nash equilibrium, optimal resource allocation and optimal bidding value for each MVNO are already shown in Section \ref{sec:pro4}.  
\begin{figure}[t]
	\centering
	\includegraphics[width=3in]{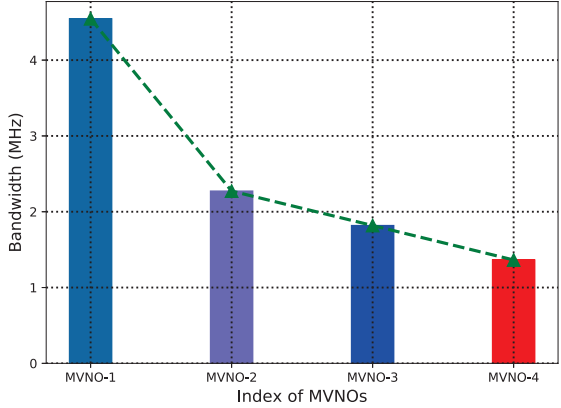}
	\caption{Bandwidth allocation to each MVNO under the proposed (GKM) algorithm.}
	\label{Sim:Bandwidth Allocation}
\end{figure}
\begin{figure*}[t]
    \centering
    \includegraphics[width=7in]{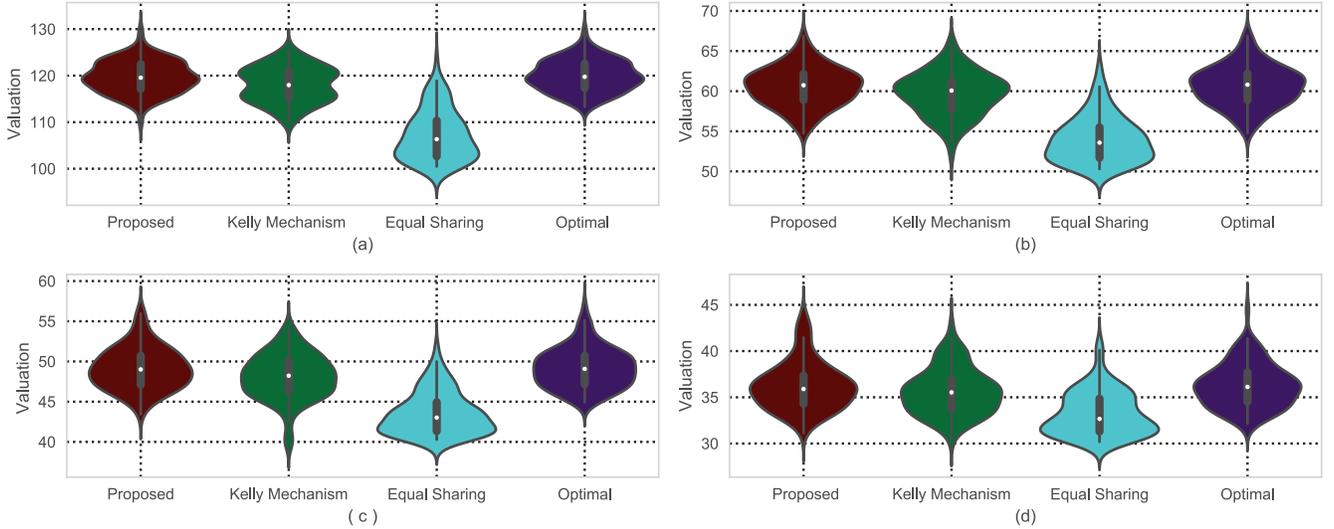}
    \caption{Comparison of the achieved valuation for (a) MVNO-1, (b) MVNO-2, (c) MVNO-3, (d) MVNO-4. }
    \label{sim:valuation}      
\end{figure*}
\begin{figure}[t]
	\centering
	\includegraphics[width=3.5in]{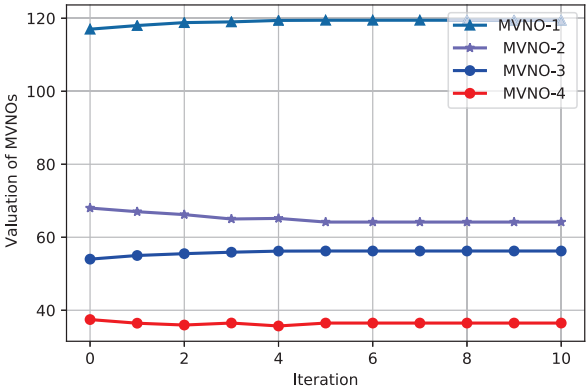}
	\caption{Convergence of valuation of MVNOs in the proposed GKM algorithm.}
	\label{fig:convergence_singler}
\end{figure}

\vspace{-0.3in} 
\section{Simulation Results}
\label{sec:pro6} 


\subsection{Simulation setting and performance metrics} In this section, we evaluate the performance of our proposed GKM algorithm in a wireless virtualized network for optimal bandwidth allocation. The network scenario of our simulation includes a single InP with one MBS and  4 MVNOs with 10, 5, 4, and 3 mobile users who are positioned randomly within the coverage area of the MBS, respectively. The radius of the macrocell is set as 500m.  At the MBS, the maximum available bandwidth is 10MHz, the maximum transmit power is 43dBm, and the thermal noise density is considered as -174dBm/Hz. The path loss model is $PL = 40\log_{10}(d_0)- 10\log_{10}(Gh_t^2h_r^2)+ 10\lambda \log_{10}(\frac{d}{d_0})+ X_g$, where $ d$  is the actual and $d_0$ is the reference distance between the transmitter and the receiver, respectively, $h_t$ and $h_r$ are respective heights of the transmitter and the receiver, and a Gaussian random variable $X_g$. The small-scale fading model is Rayleigh fading.

In this work, we maximize both aggregate valuation of MVNOs, and the individual valuation of each MVNO by optimizing the performance metrics of both bandwidth and power.
\color{black}
\subsection{Detailed Numerical Results}
In this section, we will discuss the detailed numerical results to show the efficacy of our proposed mechanism in a wireless virtualized network.

Fig.~\ref{Sim:Bandwidth Allocation} shows the bandwidth resource  allocated to the individual MVNOs under the proposed GKM algorithm. Here, MVNO-1 is allocated 4.5455MHz, MVNO-2 is allocated 2.273MHz, MVNO-3 is 1.812MHz and MVNO-4 is 1.364MHz, respectively. From the above results, we observe that the MVNO who has more mobile users receives a larger fraction of bandwidth resources owned by the InP. For this reason, in Fig.~\ref{Sim:Bandwidth Allocation} MVNO-1 is allocated more bandwidth resources compared with the others. It is also clear that the MVNO with more mobile users will invest or bid much more than other MVNOs to get more bandwidth to fulfill the service requirement of its mobile users.
\begin{figure*}[t!]
	\centering
	\includegraphics[width=7in]{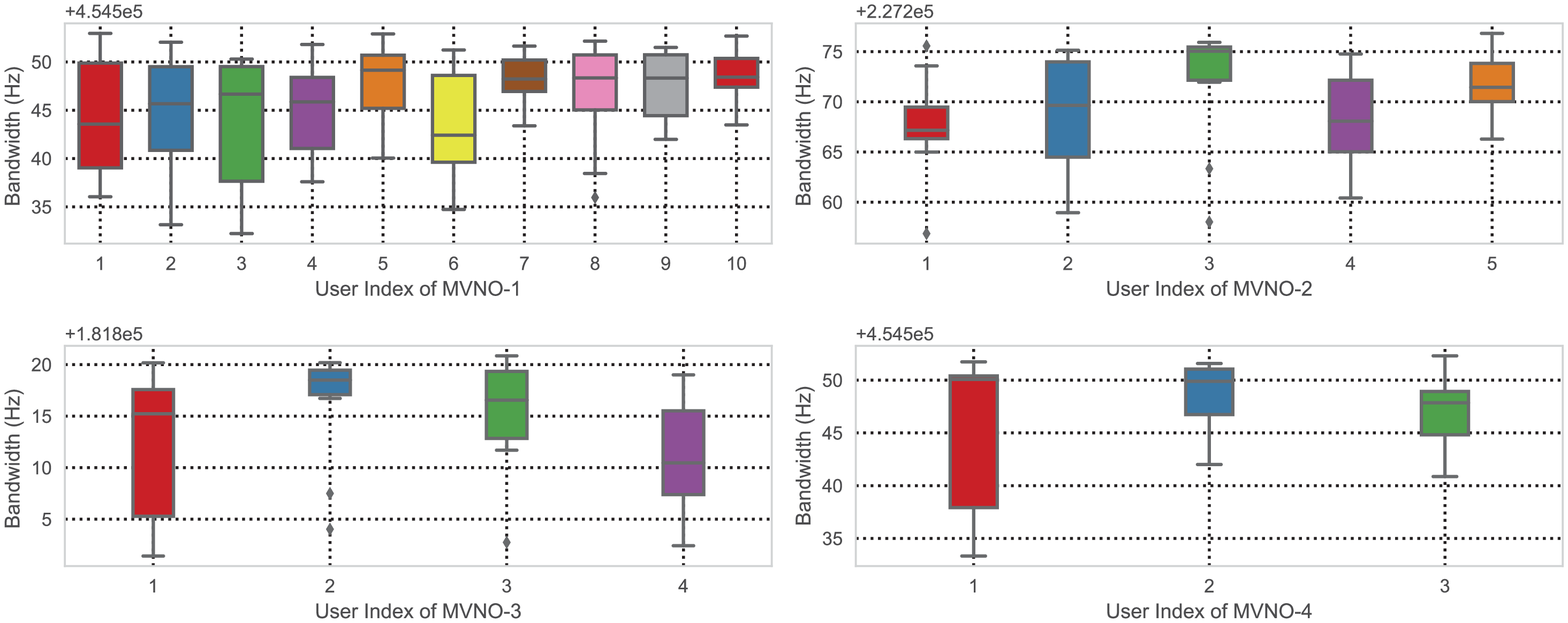}
	\caption{ Bandwidth allocation to each user of (a) MVNO-1, (b) MVNO-2, (c) MVNO-3, (d) MVNO-4. }
	\label{sim:Bandwidth Allocation MVNO Users}
\end{figure*}  

 Fig.~\ref{sim:valuation} compares the achieved valuation for each MVNO as the function of allocated bandwidth resource under different algorithms: our proposed algorithm, the traditional Kelly Mechanism \cite{kelly1997charging}, the Equal Sharing and the Optimal solutions. Under the Equal Sharing mechanism, the InP allocates an equal amount of bandwidth to all MVNOs. The Optimal solution is achieved under zero market influence power of MVNOs, i.e., no MVNO can alter the market price of the resource. This scenario accounts for the price-taking buyers (MVNOs) which is of our particular interest. As an example, from Fig.~\ref{sim:valuation}, we observe that the median of the achieved valuation of MVNO-1 is  119 (Proposed), 117 (Kelly Mechanism), 106 (Equal Sharing), and 120 (Optimal). Moreover, we can also see the lowest and the highest valuation of MVNO-1 is 106 - 133.5 (Proposed), 105.4 - 130 (Kelly Mechanism), 93 - 129.3 (Equal Sharing), 109 - 133.6 (Optimal). Therefore, our proposed algorithm achieves a higher valuation than doing the traditional Kelly Mechanism and Equal Sharing for all MVNOs. Also, it is comparatively close to the Optimal solution, demonstrating its efficacy. It is clear that our solution approach is better than the traditional Kelly Mechanism and the Equal Sharing.
 
  Furthermore, Fig. \ref{fig:convergence_singler} demonstrates the convergence of the achievable valuation of MVNOs in the proposed Generalized Kelly Mechanism (GKM). At the beginning of the algorithm, as the fraction of bandwidth is randomly allocated to each MVNO, the MVNOs evaluate the valuation randomly. In the subsequent iterations, each MVNO chooses its best strategy, i.e., bidding value, to achieve the highest valuation. From Fig. \ref{fig:convergence_singler}, we observe that our proposed GKM algorithm converges to the equilibrium point in just 5 iterations. Here, MVNO-1 achieves the highest valuation compared with the other MVNOs. This is because it has more associated users, and gets more fraction of bandwidth from the InP.
\begin{figure}[t!]
	\centering
	\includegraphics[width=3in]{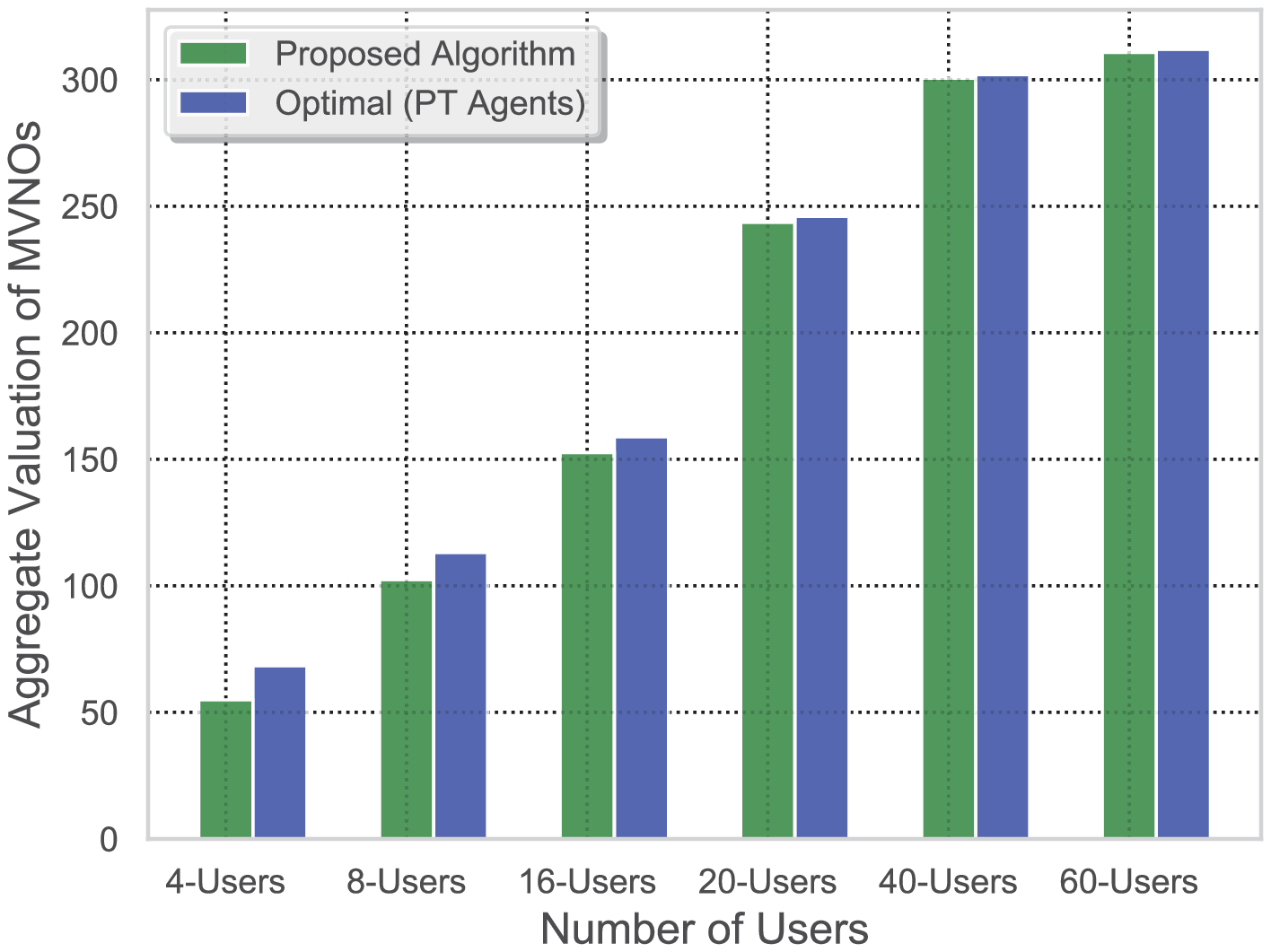}
	\caption{Aggregate valuation of MVNOs for different number of users.}
	\label{sim:agg_valuation}
\end{figure}
\begin{figure}[t!]
	\centering
	\includegraphics[width=3in]{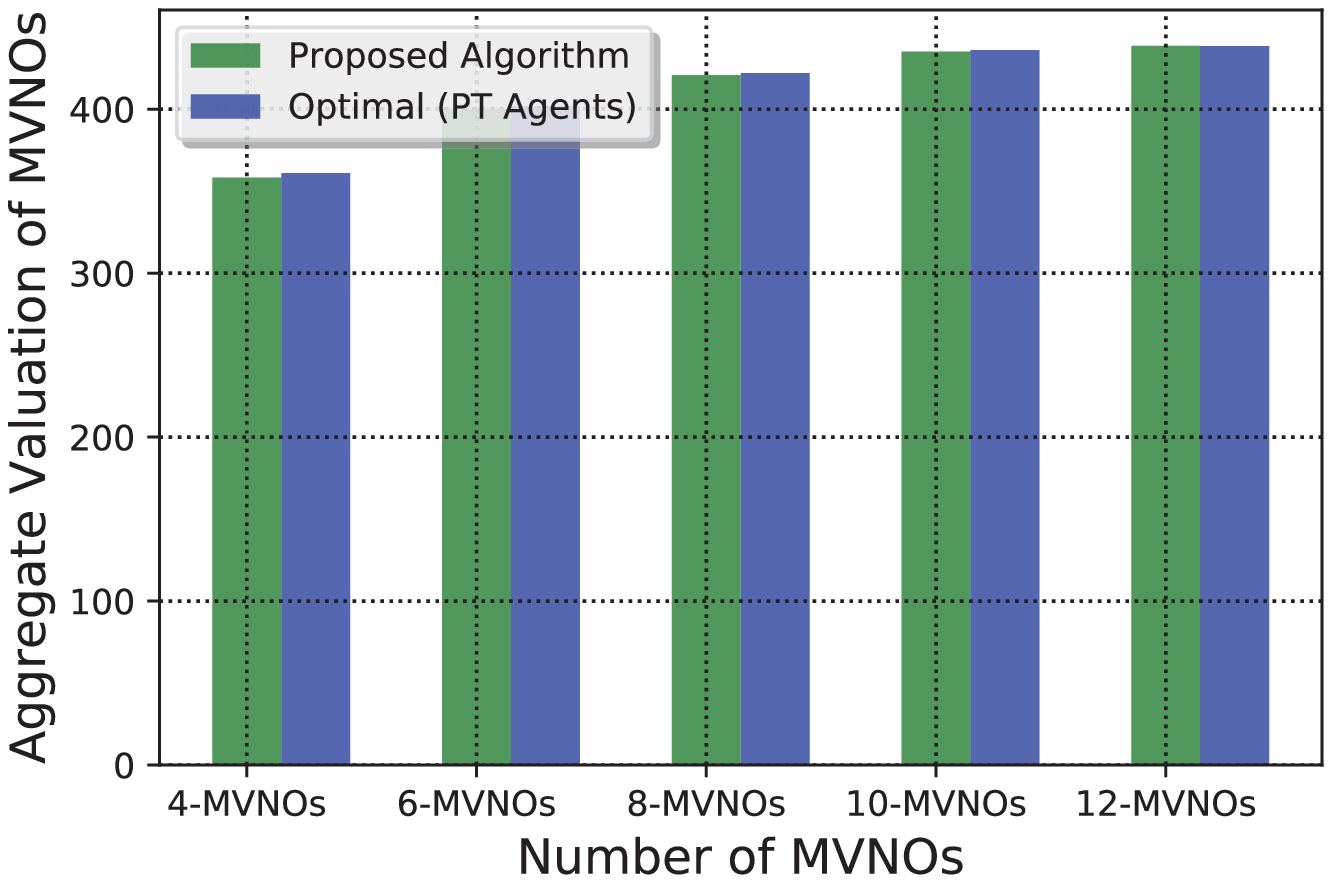}
	\caption{Aggregate valuation of MVNOs under different number of MVNOs.}
	\label{fig:aggmvno}
\end{figure} 

 Fig.~\ref{sim:Bandwidth Allocation MVNO Users} demonstrates the solution for the lower-level problem. The bandwidth resource for individual users in each MVNO is assigned as per \eqref{eq:resource_alloc}. From Fig.~\ref{sim:Bandwidth Allocation MVNO Users}, we can notice that the amount of bandwidth that MVNOs allocated to each of their users. As an example, for MVNO-1, the median of the allocated bandwidth to its mobile users are  around (0.44, 0.46, 0.47, 0.46, 0.49, 0.43, 0.48, 0.48, 0.48, 0.49) MHz. From results, the amount of bandwidth allocation among users of the same MVNO are different.

In Fig.~\ref{sim:agg_valuation}, we show the adversity of the number of users in the network. We observe that our proposed algorithm obtains the Optimal solution for the larger network. This is due to the increase in the bidding value of the corresponding MVNOs to the users for obtaining resources. Consequently, with the increase in the number of buyers in the network, the market price is less likely to be affected as discussed before. Similarly, with the increase in the number of MVNOs, the achievable aggregate valuation of each MVNO will be the same as the optimal social welfare, as observed in Fig. \ref{fig:aggmvno}.

\begin{figure*}[t]
	\centering
	\includegraphics[width=3.9in]{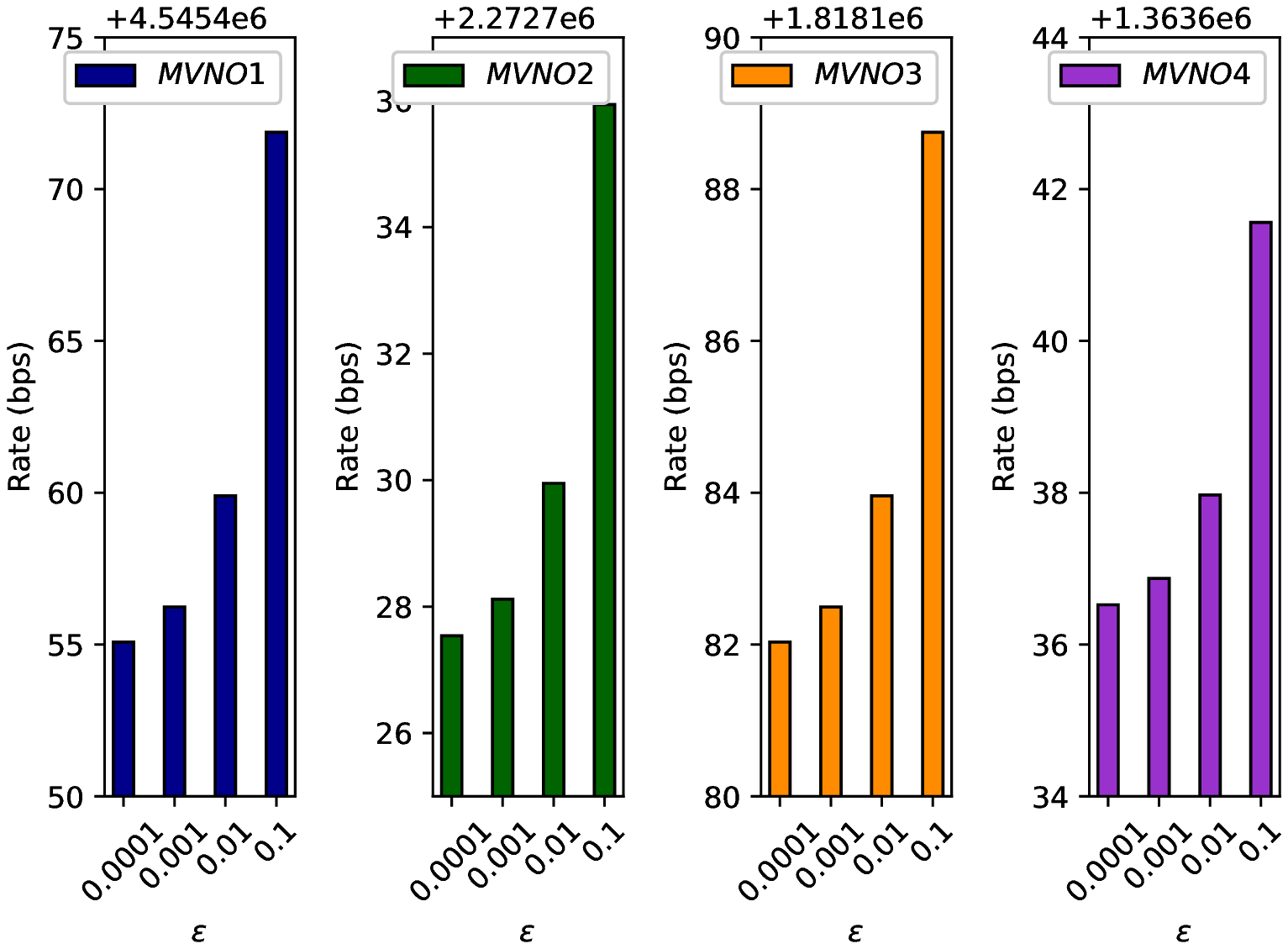}
	\caption{Rate achieved by each MVNO with respect to different outage thresholds.}
	\label{sim:outage_threshold}
\end{figure*}

Fig.~\ref{sim:outage_threshold} depicts the achieved data rate for each MVNO under incomplete information. It is evident that the data rate increases with a more relaxed outage constraint threshold. 
For a sufficiently large outage threshold, a significant gain in the achieved data rate is observed.

\begin{figure}[t]
	\centering
	\includegraphics[width=3in]{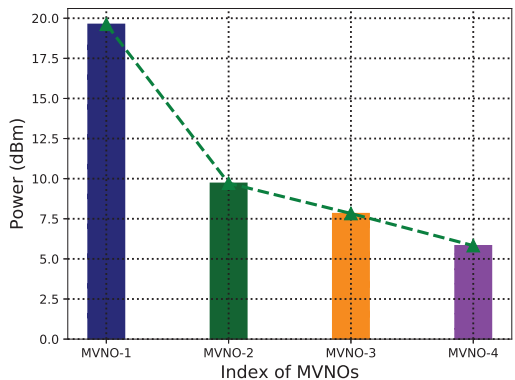}
	\caption{Power allocation at MVNOs.}
	\label{sim:power_MVNOs}
\end{figure}

\begin{figure*}[t]
	\centering
	\includegraphics[width=7in]{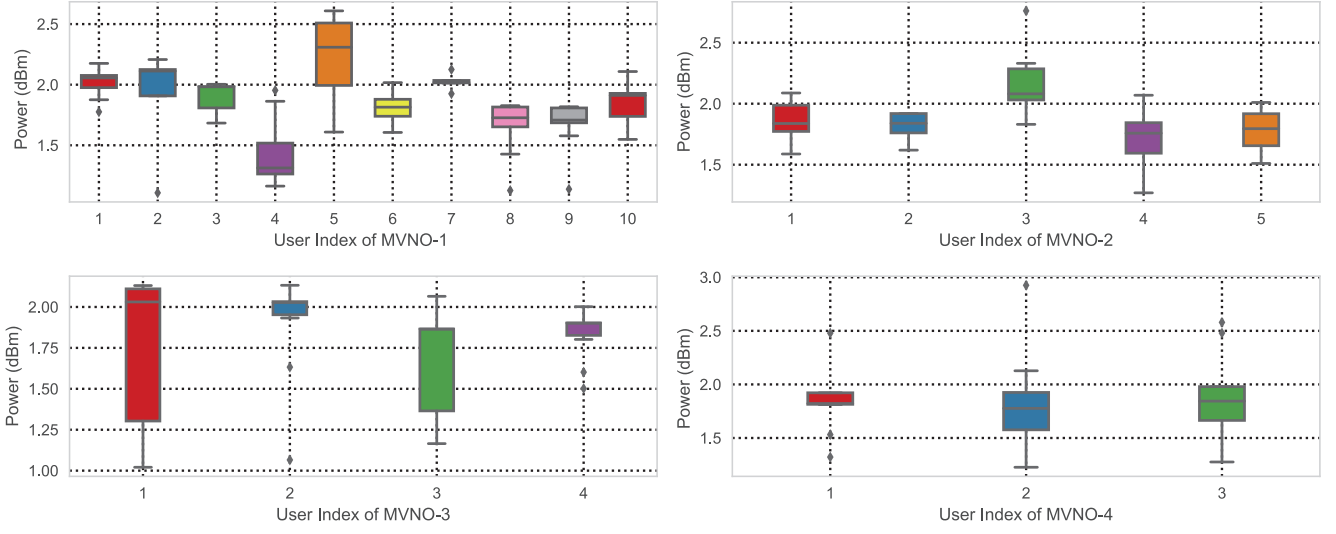}
	\caption{ Power allocation to each user of (a) MVNO-1, (b) MVNO-2, (c) MVNO-3, (d) MVNO-4.}
	\label{sim:Power_users}
\end{figure*}

Fig.~\ref{sim:power_MVNOs} represents the allocated power to each MVNO under our proposed algorithm where the InP allocates 19.625dBm to MVNO-1, 9.717dBm to MVNO-2, 7.83dBm to MVNO-3 and 5.827dBm to MVNO-4, respectively. This result is similar to the bandwidth allocation because the MVNO who has more mobile users gets a larger share of the resources. We also show the power allocation from each MVNO to its respective users in Fig.~\ref{sim:Power_users}, which is the solution of the lower-level problem. As we have discussed in the lower-level of the bandwidth allocation problem, the power allocation to each user depends on the channel condition which is assigned to that user.   

\begin{figure*}[t]
	\centering
	\includegraphics[width=7in]{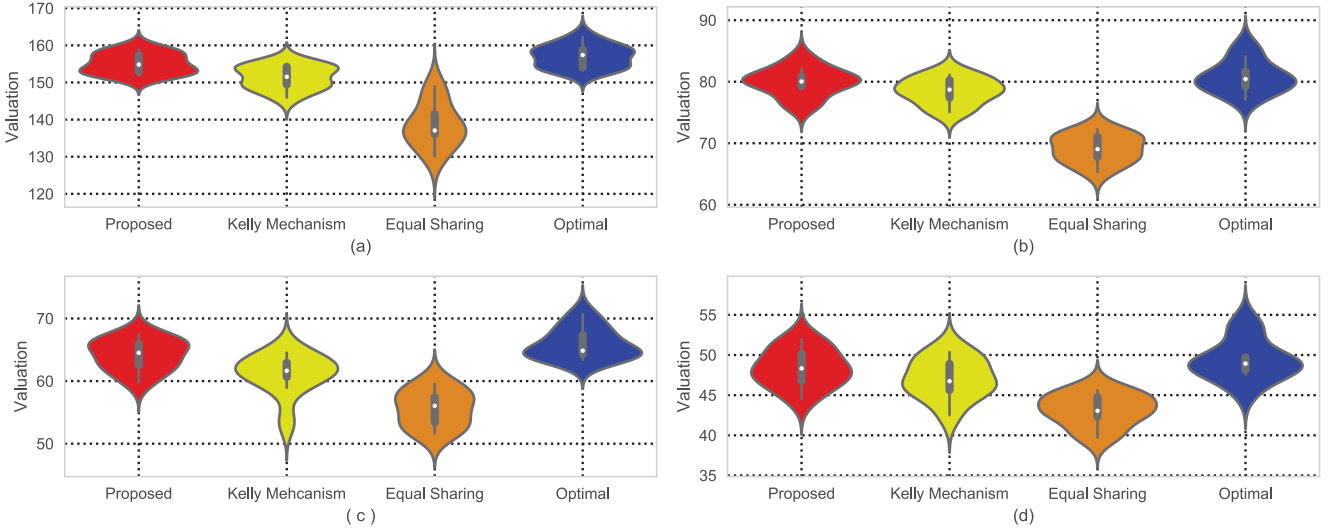}
	\caption{Comparison of achieved valuation under multiple resources for (a) MVNO-1, (b) MVNO-2, (c) MVNO-3, (d) MVNO-4.}  
	\label{sim:agg_power_bw}    
\end{figure*}

\begin{figure}
	\centering
	\includegraphics[width=3in]{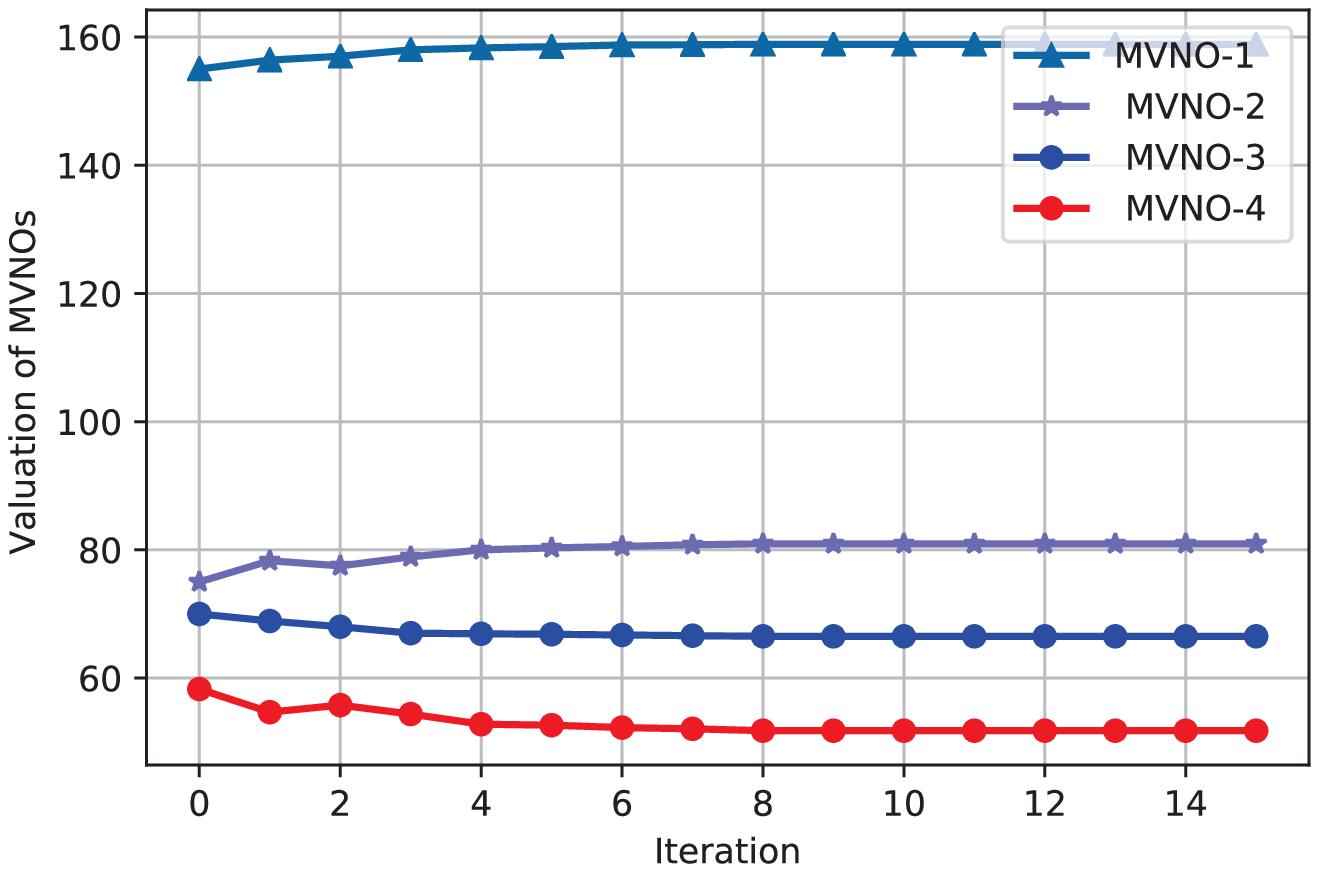}
	\caption{Convergence of valuation of MVNOs for multiple resources in the proposed GKM algorithm.}
	\label{fig:convergence_multir}
\end{figure}

Finally, Fig.~\ref{sim:agg_power_bw} shows the achieved valuation for each MVNO  as the function of allocated bandwidth and power resources by the InP under different algorithms. Similar to the individual resource allocation, our proposed algorithm results in a higher valuation than do the traditional Kelly Mechanism and Equal Sharing scheme. Further, we observe that the valuation is comparatively near to the Optimal solution. In Fig. \ref{fig:convergence_multir}, we present the convergence of valuations of all MVNOs under multiple resources (i.e., power, and bandwidth) allocation. We observe the convergence of our proposed algorithm in lesser than 8 iterations.

\section{Conclusion}
\label{sec:pro7}
In this paper, we have formulated a two-level optimal bandwidth allocation problem for wireless network slicing. In the upper-level, the GKM is introduced to model MVNOs as bidders who compete for the bandwidth from the InP in order to serve their mobile users. The InP, who is the seller of the resources, then executes the bandwidth allocation process under the GKM to fulfill these requests. In the lower-level, each MVNO allocates the optimal bandwidth resource to its users. Morever, we consider the incomplete information scenario where the MVNO does not know the exact channel state information for its users. Finally, we consider the multiple resource scenario where  MVNOs compete with each other for power and bandwidth resources from the InP. Simulation results have reflected that the aggregated valuation of the MVNOs following our proposed algorithm outperforms that by the traditional Kelly Mechanism, Equal Sharing, and is nearly close to the Optimal. 

\begin{appendices}  
\section{Proof of Proposition 1} 
Differentiating (6) w.r.t the bidding value $b_m$, the stationary condition of (6) can be obtained as 
\begin{equation}
v'(r_m(\textbf{b}))\frac{\partial r_m(\textbf{b})}{\partial b_m} - q_m = 0.     \tag{40}
\end{equation}
MVNO $m \in \mathcal{M} $ in this resource competition is a price-anticipating agent, and the resource  $r_m(\textbf{b})$ allocated to MVNO $m$ is dependent on its bidding value $b_m$. Therefore, using (10), $r_m(\textbf{b})= \frac{b_m}{\beta}$, and applying the first-order derivative, we have 
\begin{equation}
 \frac{\partial r_m(\textbf{b})}{\partial b_m} = \frac{1}{\beta} \left(1 - \frac{b_m}{\beta} \frac{\partial \beta}{\partial b_m} \right).   \tag{41} 
 \end{equation} 
 
By using (10), the above (40) and (41) are rewritten as 
\begin{equation}
v'(r_m(\textbf{b}))\left(1- \frac{b_m}{\beta}\right) = \beta q_m.   \tag{42} 
\end{equation}
According to (10), the optimal bidding strategy of MVNO $m$ is 
\begin{equation}
b_m = \frac{1}{q_m}r_m(\textbf{b}) v'(r_m(\textbf{b}))(1- \mu_m), \ \forall m \in \mathcal{M}.  \tag{43} 
\end{equation}
\section{Proof of Proposition 2}
We will prove that the penalty value for each MVNO $m \in \mathcal{M}$ depends on its bidding value.       
	  
	  \begin{align*}
	  \frac{\partial u_m}{\partial b_m}
	       = v'_m(r(\textbf{b}))\bigg(\frac{(\sum_{m=1}^{M}b_m-b_m) R}{(\sum_{m=1}^{M}b_m)^2}\bigg)-q_m &=0,  \tag{44}\\
	       \Leftrightarrow \frac{1}{\beta}v'_m(r_m(\textbf{b}))(1- \frac{r_m(\textbf{b})}{R})- q_m &= 0,  \tag{45} \\
	       \Leftrightarrow \frac{1}{\beta}v'_m(r_m(\textbf{b}))(1- \frac{r_m(\textbf{b})}{R})&= q_m \tag{46}.    
	  \end{align*}
 
\section{proof of unique penalty for each MVNO} 

From (43) and (10),
\begin{equation*}
\frac{1}{q^*_m} v'_m(r^*_m(\textbf{b}))(R- r^*_m(\textbf{b})) = \sum_{m=1}^{M}b_m , \forall m \in \mathcal{M}.   \tag{47} 
\end{equation*}
Therefore,
\begin{align*}
q^*_m : q^*_n &= v'_m(r^*_m(\textbf{b}))(R - r_m(\textbf{b})): v'_n(r^*_n(\textbf{b}))(R - r^*_n(\textbf{b})),\\
& \quad \quad \quad \quad \quad \quad \quad \quad \quad   \    \forall m,n \in \mathcal{M}.   \tag{48}
\end{align*}
Suppose the penalty vector $\textbf{q}$ induces the optimal bandwidth allocation vector $\textbf{r}$. The optimality condition for the optimal bandwidth allocation is $v'_m(r^*_m(\textbf{b})) = v'_n(r^*_n(\textbf{b})), \  \forall m,n \in \mathcal{M}$ \cite{yang2013price}. Therefore, (48) becomes
\begin{equation}
\frac{R - r^*_m(\textbf{b})}{q^*_m} = \frac{R - r^*_n(\textbf{b})}{q^*_n}= \frac{MR - R}{\sum_{m=1}^{M} q^*_m},  \forall m,n \in \mathcal{M}.   \tag{49}  
\end{equation} 
From (42),
\begin{equation*}
\frac{R - r^*_m(\textbf{b})}{M - 1} = \frac{q^*_m}{\sum_{m=1}^{M}q^*_m} R.  \tag{50} 
\end{equation*}
Inspired by the optimal condition in (50), we iteratively update the penalty of each MVNO $ m \in \mathcal{M} $ using the information of the previous iteration. Therefore, at each iteration, the InP updates the penalty of each MVNO according to
\begin{equation*}
q_m^k = q_m^{k-1} + \left(\frac{R - (r_m(\textbf{b}))^{k-1}}{M-1} - \frac{Rq_m^{k-1}}{\sum_{m=1}^{M}q_m^{k-1}}\right), 
	 \forall m \in \mathcal{M}.    \tag{51} 
\end{equation*}  
\section{Proof of Theorem 1}

We prove that a unique Nash equilibrium exists in our proposed resource competition among MVNOs with the different bidding strategies $b_m>0,\forall m\in \mathcal{M}$.

\balance First, we have
 \begin{align*}
 \frac{\partial u_m}{\partial b_m}
      = v'_m(r(\textbf{b}))\bigg(\frac{b_{-m} R}{(\sum_{m=1}^{M}b_m)^2}\bigg)-q_m &=0  \tag{52}\\
      \Leftrightarrow\frac{1}{q_m}v'_m\left(\frac{b_m R}{\sum_{m=1}^{M}b_m}\right)\left(\frac{b_{-m}R}{(\sum_{m=1}^{M}b_m)^2}\right)&=1  \tag{53}\\
      \Leftrightarrow \frac{1}{q_m}v'_m\left(\frac{b_m R}{\sum_{m=1}^{M}b_m}\right) \\ \qquad \times\left(\frac{R}{\sum_{m=1}^{M}b_m}- \frac{b_m R}{(\sum_{m=1}^{M}b_m)^2}\right) &=1.\tag{54}       
 \end{align*}
 When $b_m>0$, it is true that
 \begin{equation*}
  \frac{1}{q_m}v'_m\left(\frac{b_m R}{\sum_{m=1}^{M}b_m}\right)\left(\frac{R}{\sum_{m=1}^{M}b_m}- \frac{b_m R}{(\sum_{m=1}^{M}b_m)^2}\right) =1.\tag{55}
 \end{equation*} 
When $b_m=0$, we also have 
\begin{equation*}
\frac{1}{q_m}v'_m(0)\leq 1.   \tag{56} 
\end{equation*} 
There exists a unique Nash equilibrium \cite{ma2016efficient} if the above two conditions are satisfied.

\section{Proof of Proposition 3} 
The first-order derivative w.r.t $r_m(\textbf{b})$ of (15) is 

\begin{equation*}
\frac{\partial \hat{v}_m(r_m(\textbf{b}))}{\partial r_m(\textbf{b})} =  \frac{1}{q_m}(1- \mu_m)v'_m(r_m(\textbf{b})).   \tag{57} 
\end{equation*}

When $\mu_m$ and $r_m(\textbf{b})$ are greater than zero, $\frac{\partial \hat{v}_m(r_m(\textbf{b}))}{\partial r_m(\textbf{b})}>0$ in (57). From (57), $(1 - \mu_m)$ is strictly decreasing in $b_m$. Moreover, the allocated bandwidth $r_m$ depends on the bidding value $b_m$ of MVNO. Therefore, $r_m(\textbf{b})$ is also decreasing. Thus, $\frac{\partial \hat{v}_m(r_m(\textbf{b}))}{\partial r_m(\textbf{b})}$ is monotonically decreasing. For this reason,  $\hat{v}_m(r_m(\textbf{b}))$ is a concave function, and hence, the optimization problem (16) has a unique maximum value. 

Here, we define the Lagrangian of (16) as 
\begin{equation*}
L (r_m, \rho)  = \sum_{ m\in \mathcal{M} } \hat{v}_m(r_m(\textbf{b})) + \rho \left[R - \sum_{m=1}^{M}r_m(\textbf{b})\right],       \tag{58} 
\end{equation*}
where $\rho \geq 0 $ is the Lagrangian multiplier for the constraint (16). Therefore, the KKT conditions can be expressed by using the first-order derivative of (58) w.r.t $r_m$ and $\rho$ as
\begin{align*}
 \frac{\partial L(r_m, \rho)}{\partial r_m(\textbf{b})} &=  \frac{1}{q_m}\left[ (1- \mu_m)v'_m(r_m(\textbf{b}))\right] - \rho  \leq 0,  \\ 
  &  \qquad \text{if ${r_m}\geq 0,$}  \forall m \in \mathcal{M}, \tag{59} \\
  \frac{\partial L(r_m, \rho)}{\partial \rho}  &= R - \sum_{m=1}^{M}r_m(\textbf{b}) \geq 0, \ \ \text{if $ \rho \geq 0$},   \tag{60}  
\end{align*} 
When $ \rho > 0$,
\begin{equation*}
\frac{1}{q_m}v'_m(r_m(\textbf{b})) (1- \mu_m) = \rho.    \tag{61} 
\end{equation*}   
From (43) and (61), $\rho = \beta$. By using (10), we can clearly observe that the optimal bidding strategy in (7) is satisfied. So, the optimal resource allocation to each MVNO at the equilibrium is defined as follows: 
\begin{equation*}
r^*_m(\textbf{b})  = \frac{b_m q_m}{v'(r_m(\textbf{b}))(1- \mu_m)}, \ \    \forall m \in \mathcal{M}.   \tag{62} 
\end{equation*}
To sum up,
\begin{equation*}
\begin{cases}
   \rho = \beta,  \\
    \rho > 0,              & \text{when} \sum_{m=1}^{M}r_m(\textbf{b})= R, \\
    \rho = 0,              &  \text{when} \sum_{m=1}^{M}r_m(\textbf{b})< R,
\end{cases}   \tag{63} 
\end{equation*}

\section{Proof of Optimal Bandwidth Allocation} 

The KKT conditions can be expressed with the first-order derivative of (22) w.r.t $x_s^{m}$ and $ \lambda$ as
\begin{align*}
 \frac{\partial L}{\partial {x_s^{m}}} &= \frac{\log_2\left(1+\frac{p_sh_s}{N_0}\right)}{x_s^m r_m\log_2\left(1+\frac{p_sh_s}{N_0}\right) + 1 }-\lambda\leq 0, \\ 
  &  \qquad \text{if ${x_s^{m}}\geq 0,$}  \forall s \in \mathcal{S}_m, \tag{64} \\
  \frac{\partial L}{\partial \lambda} &= 1-\sum_{s=1}^{S_m}x_s^m \geq 0, \ \ \text{if $ \lambda\geq 0$},  \tag{65}  
\end{align*} 
Solving (64) gives the bandwidth allocated to each user $s \in \mathcal{S}_m$ as
\begin{equation*}
x_s^{m*} = \frac{1}{\lambda^{*} r_m} -  \frac{1}{r_m \log_2\left(1+\frac{p_sh_s}{N_0}\right)},  \ \ \ \forall s \in \mathcal{S}_m,  \tag{66} 
\end{equation*}
where
\begin{equation*}
\frac{1}{\lambda^*} = \frac{1}{|S_m|} \left[r_m + \sum_{s=1}^{S_m} \frac{1}{\log_2\left(1+\frac{p_sh_s}{N_0}\right)} \right]^+.  \tag{67} 
\end{equation*}
Thus, the optimal bandwidth allocated to each user $s \in \mathcal{S}_m$ is 
\begin{equation*}
x_s^{m*} = \frac{1}{r_m} \left( \frac{1}{|S_m|} \left[r_m + \sum_{s=1}^{S_m} \frac{1}{\alpha^{*}} \right] - \frac{1}{\alpha^{*}} \right) , \forall s \in \mathcal{S}_m     \tag{68} 
\end{equation*}
where $\alpha^*= \log_2\left(1+\frac{p_sh_s}{N_0}\right)$.

\end{appendices} 
 
%
\IEEEpeerreviewmaketitle

\ifCLASSOPTIONcaptionsoff
  \newpage
\fi

\bibliographystyle{IEEEtran}
\end{document}